\documentclass[12pt, preprint]{aastex}

\newcommand{\Un}{U_n}
\newcommand{\R}{{\cal R}}
\def\expec#1{\left\langle#1\right\rangle}

\newcommand{\Ldenv}{{\cal L}_V}

\newcommand{\Mstarv}{M^{\ast}_V}

\newcommand{\Phistar}{\Phi^{\ast}}
\newcommand{\lbox}{\mathrm{L_{box}}}

\newcommand{\lam}{\lambda}
\newcommand{\Lam}{\Lambda}

\newcommand{\hinv}{{h^{-1}}}
\newcommand{\mpc}{{\rm\,Mpc}}
\newcommand{\kpc}{{\rm\,kpc}}
\newcommand{\himpc}{\hinv{\rm\,Mpc}}

\newcommand{\yr}{{\rm yr}}
\newcommand{\Msun}{M_{\odot}}

\newcommand{\Lsunv}{L_{\odot,V}}

\newcommand{\Mstellar}{\rm\,M_{\rm stellar}}
\newcommand{\Zsun}{Z_{\odot}}
\newcommand{\himsun}{\hinv{\Msun}}

\newcommand{\etal}{et~al.}

\newcommand{\ltsim}{\lesssim}
\newcommand{\gtsim}{\gtrsim}
\def\bgeqa#1{\begin{eqnarray}\label{#1}}
\def\endeqa{\end{eqnarray}}

\def\Fig#1{Figure~\ref{#1}}

\shorttitle{The Lyman Break Galaxies: progenitors \& descendants}
\shortauthors{Nagamine}

\begin{document}

\title{The Lyman Break Galaxies: their progenitors and descendants}


\author{Kentaro Nagamine}
\affil{Joseph Henry Laboratories, Physics Department, Princeton University,\\ 
Princeton, NJ 08544, USA}
\email{nagamine@astro.princeton.edu}


\begin{abstract}
We study the evolution of Lyman Break Galaxies (LBGs) from $z=5$ 
to $z=0$ by tracing the merger trees of galaxies in a large-scale 
hydrodynamic simulation based on a $\Lam$ cold dark matter model.  
In particular, we emphasize on the range of properties 
of the sample selected by the rest-frame $V$ band luminosity,
in accordance with recent near-IR observations.
The predicted rest-frame $V$ band luminosity function agrees
well with the observed one when dust extinction is taken into account.
The stellar content and the star formation histories of LBGs
are also studied. We find that the LBGs intrinsically brighter than 
$M_V=-21.0$ at $z=3$ have stellar masses of at least $10^9\Msun$,
with a median of $10^{10}\himsun$. The brightest LBGs ($M_V\ltsim -23$) 
at $z=3$ merge into clusters/groups of galaxies at $z=0$, as suggested 
from clustering studies of LBGs. Roughly one half of the galaxies
with $-23\ltsim M_V \ltsim -22$ at $z=3$ fall into groups/clusters, 
and the other half become typical $L^*$ galaxies at $z=0$ with stellar 
mass of $\approx 10^{11}\Msun$. Descendants of LBGs at the present
epoch have formed roughly 30\% of their stellar mass by $z=3$,
and the half of their current stellar population is 10 Gyr old,
favoring the scenario that LBGs are the precursors of the present
day spheroids.
We find that the most luminous LBGs have experienced a starburst
within 500 Myr prior to $z=3$, but also have formed stars continuously
over a period of 1 Gyr prior to $z=3$ when all the star formation
in progenitors is coadded.
We also study the evolution of the mean stellar metallicity 
distribution of galaxies, and find that the entire distribution 
shifts to lower metallicity at higher redshift. 
The observed sub-solar metallicity of LBGs at $z=3$ is naturally 
predicted in our simulation.
\end{abstract}


\clearpage
\section{Introduction}
\label{section:intro_lbg}

A large sample of galaxies at $z=3-4$ has become available
owing to the observational technique of selecting high-redshift
galaxies by broad-band color signatures when the Lyman limit 
spectral discontinuity at 912~\AA ~passes through different filters; 
the so-called Lyman Break Galaxies \citep[LBGs;][]{Steidel92, Steidel95}.
So far, most surveys have been carried out at optical 
wavelengths, which corresponds to LBGs in the rest-frame far ultra-violet 
(UV). Such observations have detected strong clustering
of LBGs \citep{Adelberger98, Giavalisco98, Steidel98}, which has 
generally been interpreted as indirect evidence that LBGs
reside in massive dark matter halos.  

However, the true character of LBGs and their evolutionary history 
remain uncertain. In one scenario, LBGs are low-mass, merger-induced 
starbursting systems, and are the precursors of present day low-mass 
spheroids \citep*{Lowenthal97, Sawicki98, Somerville01}. 
In another scenario, they have stellar mass of $\approx 10^{10}\Msun$, 
sitting in massive dark matter halos and continuously forming stars 
over a period of 1 Gyr. They evolve into bright elliptical and spiral 
galaxies at $z=0$, and later merge into clusters or groups of galaxies 
\citep{Mo96, Steidel96, Baugh98}. Another possibility is that they are 
massive galaxies experiencing merger-induced starbursts \citep{Somerville01}. 

In order to clarify the stellar content and the star formation 
histories of LBGs, near infra-red (IR) studies of LBGs have recently 
been carried out \citep*{Sawicki98, Papovich01, Pettini01, Shapley01, 
Rudnick01}. 
These observations directly probe the rest-frame optical properties  
of LBGs at $z=3$; wavelengths less affected by dust obscuration. 
However, the number of galaxies used in these studies is still small,
and the inferred star formation time-scale and the values of extinction 
differ somewhat among authors due to different sample selection and 
different modeling of SEDs used in their analyses.
Nevertheless, the median value of extinction seems to lie in the range
$E(B-V)\approx 0.1-0.3$, and the recent studies seem to indicate
the existence of a significant old stellar population in LBGs at $z=3$
with the stellar mass of $\approx 10^{10}\Msun$ 
\citep{Papovich01, Shapley01}. 
In addition, \citet{Pettini01} used the observed oxygen lines 
in the rest-frame optical to estimate the metallicity of LBGs to be 
$0.1-1.0\Zsun$.

In this article, we focus on the merger history and
the star formation history of LBGs from $z=5$ to 0,
as well as their stellar content.
We follow the merger trees of galaxies both backward and 
forward in time, and examine the stellar mass of progenitors and 
descendants of LBGs as well as their star formation histories.
We hope to provide some insights on how LBGs fit into the overall 
galaxy formation picture in a cold dark matter (CDM) model; 
in particular, we address what kind of objects LBGs form from, 
and what type of objects LBGs evolve into.

In our earlier papers \citep*[][hereafter Paper I and II]
{Nagamine01a, Nagamine01b}, we discussed the star formation history, 
stellar metallicity distribution, luminosity function, and color 
distribution of galaxies in a $\Lam$CDM universe using the same
simulation.
There we showed that the agreement between local observations 
and the simulation was relatively good, considering the 
uncertainties involved on both sides, and argued that we are beginning 
to obtain meaningful results for the global properties of galaxies 
in large-scale hydrodynamic simulations, with the hydrodynamic mesh 
approaching $\sim (1000)^3$. The simulation used in this paper has 
a box size of $\lbox=25\himpc$ with Eulerian hydrodynamic mesh 
of $768^3$ and cosmological parameters of $(\Omega_m, \Omega_{\Lam}, 
\Omega_b, h, \sigma_8)=(0.3, 0.7, 0.035, 0.67, 0.9)$ 
(See Paper~I for the details of the simulation).

Other numerical studies of LBGs using hydrodynamic simulations exist.
\citet{Dave99} and \citet*{Weinberg00} used a Smoothed Particle 
Hydrodynamic (SPH) simulation with a box size 
of $11.1\himpc$, which was stopped at $z=2$, due to a lack of long 
wavelength perturbations. Therefore, they cannot check 
whether the properties of the simulated galaxies are 
consistent with local galaxy observations, in particular, 
the luminosity function, the color distribution of galaxies, 
and the stellar matter density at $z=0$. 
On the other hand, the simulation used in this paper is continued
up to $z=0$.
The advantage of SPH simulations is much higher spatial resolution
than that of Eulerian simulations, but the SPH mass resolution is 
usually about an order of magnitude worse than that of the Eulerian 
method which is adopted in this paper.
With these differences in the simulations, it is interesting to 
compare the results from the two types of simulations.
Comparisons with the work of \citet{Dave99} and \citet{Weinberg00} 
will be discussed in the following sections. The present work 
extends and complements earlier works. 

As in Paper~II, we identify galaxies using the grouping algorithm
HOP \citep{Eisenstein98}, and use the isochrone synthesis model GISSEL99
\citep[][Charlot 1999, private communication]{BC93} to obtain 
the stellar luminosity output in the simulation.  
We assume an initial mass function of \citet{Salpeter55} with 
a turnover at the low-mass end as reported by \citet{Gould96}. 

In \S~\ref{section:selection_lbg}, we describe how we select 
LBGs in our simulation.
The rest-frame $V$ band luminosity function of galaxies at $z=3$ 
in the simulation is presented in \S~\ref{section:lf_lbg}, 
which is compared with the observed one. 
We discuss the stellar mass of LBGs in \S~\ref{section:stellarmass_lbg}.
The merger trees and the star formation histories of LBGs are
presented in \S~\ref{section:mergertree_lbg} ($z\geq 3$) and
\S~\ref{section:fate_lbg} ($z\leq 3$), and the progenitor
masses and the fate of LBGs are discussed. 
The star formation rate of LBGs is studied in \S~\ref{section:sfr_lbg}, 
with emphasis on instantaneous vs. continuous star formation. 
In \S~\ref{section:metal_lbg}, we discuss the metallicity
of LBGs and the evolution of the entire luminosity-metallicity 
relation. We conclude in \S~\ref{section:conclusion_lbg}. 


\section{Selection of LBGs}
\label{section:selection_lbg}

Observers select galaxies efficiently at $z\sim 3$ by using a
color selection criteria in color-color space 
\citep[e.g. with $\Un, G, \R$ filters;][]{Steidel99}.
By doing so, they select out star forming galaxies that are bright
in far-UV at $z\sim 3$.
Therefore, it is often assumed that LBGs have similar properties
to the local starburst galaxies. Based on this expectation, 
many authors use an empirical extinction law of local starburst 
galaxies \citep[e.g.][]{Calzetti97} to model the internal 
dust extinction in these galaxies.

\begin{figure}
\epsscale{1.0}
\plotone{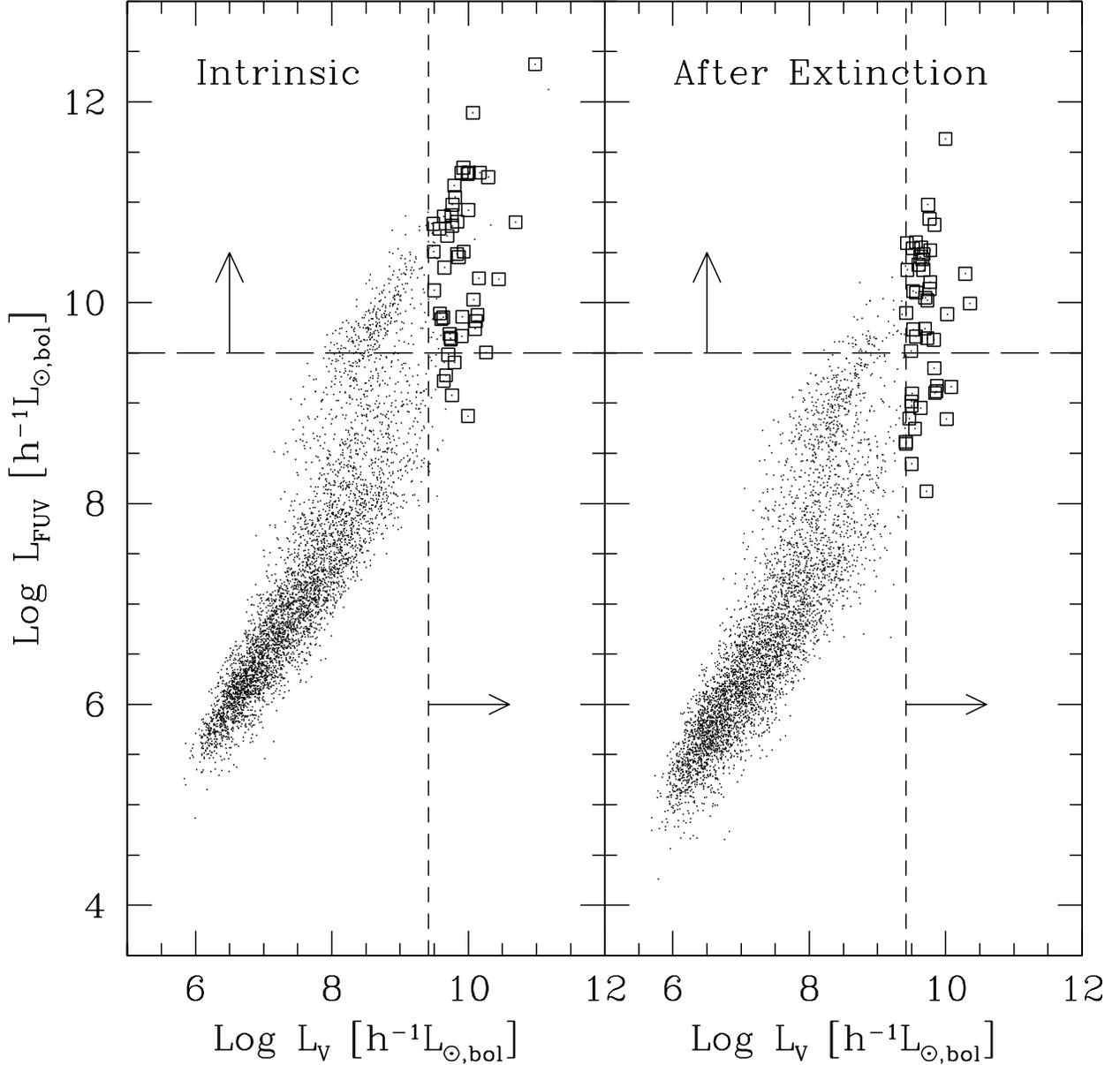}
\caption{
Rest-frame $V$ band luminosity vs. rest-frame far-UV luminosity 
(summing all luminosity below 1700~\AA) of galaxies at $z=3$ 
in the simulation. 
The left panel shows the intrinsic luminosity (before dust extinction), 
and the right panel is after a dust extinction model is applied. 
The magnitude limit ($K_s=22.5$) of a near-IR observation 
\citep{Shapley01} is indicated by the vertical short-dashed line and 
the arrow toward right. Open squares indicate galaxies that are brighter
than this magnitude limit after dust extinction. 
The horizontal long-dashed line and the upward arrow roughly indicate 
the far-UV selection criteria of $\R<25.5$ \citep{Shapley01}. 
Note that the rest-frame $V$ band selected sample takes a wider range of 
values in far-UV luminosity. 
\label{Vlum_fuvlum.eps}}
\end{figure}

If we were to mimic the observational color-selection criteria,
we would need to obtain the extinction $E(B-V)$ for each galaxy with 
some assumptions; for example, its proportionality to the metallicity 
and the star formation rate of each galaxy.
An empirical extinction law for the wavelength 
dependence (e.g. Galactic, Small Magellanic Cloud, or starburst) 
also has to be assumed. 
The position of galaxies in a color-color space 
depends significantly on the effect of dust extinction,
which is highly uncertain, 
and therefore the selection of galaxies will depend on the assumptions 
made for the calculation of $E(B-V)$ and the assumed extinction law. 
In particular, galaxies which have very high star 
formation rate may be strongly obscured and do not satisfy 
the color-selection criteria in the model. However, 
in reality, it is not clear how strongly these star-forming 
galaxies are obscured (e.g. there may be less obscuration due to the 
sweeping of dust clouds by galactic winds \citep{Pettini01, Aguirre01}.
Therefore, those galaxies which do not pass the color-selection criteria
after dust extinction in the simulation may well be detected 
in reality within the current magnitude limit in the rest-frame $V$ 
band given that they are also intrinsically bright in the optical 
wavelengths as we show in \Fig{Vlum_fuvlum.eps}.

In the left panel of \Fig{Vlum_fuvlum.eps}, we plot 
the computed rest-frame $V$ band luminosity and the rest-frame far-UV
luminosity (summing all flux below $1700$\AA) of galaxies
at $z=3$ in our simulation before taking dust extinction into account
(i.e. the intrinsic luminosity). 
The magnitude limit of $M_V<-21.0$ (which roughly corresponds to 
$K_s=22.5$ of the near-IR observations by \citet{Shapley01} with $h=0.67$) 
is indicated by the vertical short-dashed line and the arrow to the right. 
The horizontal long-dashed line and the upward
arrow roughly indicate the far-UV selection criteria of
$\R<25.5$ \citep{Shapley01}. 
Note that the rest-frame $V$ band selected sample includes a broad 
range of values of far-UV luminosity.
Most current near-IR samples are first selected in the rest-frame far-UV, 
and then in the near-IR, therefore they are drawn from the upper right 
corner of this diagram \citep{Sawicki98, Papovich01, Shapley01}.
However, near-IR selected samples are becoming available \citep{Rudnick01},
and there will be more in the future.  

\notetoeditor{Place \Fig{Vlum_fuvlum.eps} here.}

Therefore, rather than introducing further uncertainties 
in our analysis by invoking a certain color-selection criteria, 
here we consider all galaxies that are brighter than $M_V=-21.0$,
and investigate the range of properties of the luminous galaxies
at $z=3$. But we do estimate the effect of dust extinction by making 
a few assumptions.
In the right panel of \Fig{Vlum_fuvlum.eps}, we show the 
same luminosity distribution after applying a dust extinction model
as follows: (i) The value of $E(B-V)$ is proportional 
to the star formation rate and the metallicity of each galaxy: 
$E(B-V)= \beta_{dust}(SFR/\expec{SFR})(Z/Z_{max})^p$.
The values of $\beta_{dust}$, $\expec{SFR}$, and $p$ are
adjusted so that the median $E(B-V)$ is 0.15, as suggested
by \citet{Shapley01}\footnote{To avoid over-extincting high SFR galaxies, 
we adopt following sets of values after fixing $p=0.2$ and 
$Z_{max}=0.75$: $\beta_{dust}=(1.5, 1.5, 0.2, 0.2)$ 
and $\expec{SFR}=(790, 200, 50, SFR)$ for galaxies with 
$SFR[\Msun/\yr]=(>200, 200-50, 50-10, 10-0)$. The SFRs used here 
were calculated with the time interval of 200 Myr. The median extinction 
value resulting from this prescription is $E(B-V)=0.15$.} 
(ii) The \citet{Calzetti00} extinction law $k(\lam)$ is assumed, 
where $F_{\lam,{\rm obscured}}=F_{\lam,{\rm intrinsic}} 
10^{-0.4 k(\lam)E(B-V)}$.
The function $k(\lam)$ is obtained empirically from local 
starburst galaxies, and the function $F_{\lam}$ is the spectral 
energy distribution of galaxies.

The decrease in the $V$ band luminosity is smaller than that in
the far-UV luminosity, as the extinction law $k(\lam)$ is
a decreasing function of wavelength. 
Galaxies that are brighter than $M_V=-21.0$ after 
dust extinction are indicated by the empty squares in 
the right panel of \Fig{Vlum_fuvlum.eps}. 
The same galaxies are indicated by the same symbol in the 
left panel. The scatter of the points is somewhat larger in the 
right panel, and the most far-UV luminous galaxies in the left panel 
have moved to lower far-UV luminosity in the right panel because 
of the assumed proportionality between $E(B-V)$ and the star 
formation rate.


\section{Rest-frame $V$ Band Luminosity Function of LBGs at $z=3$}
\label{section:lf_lbg}

\begin{figure}
\epsscale{1.0}
\plotone{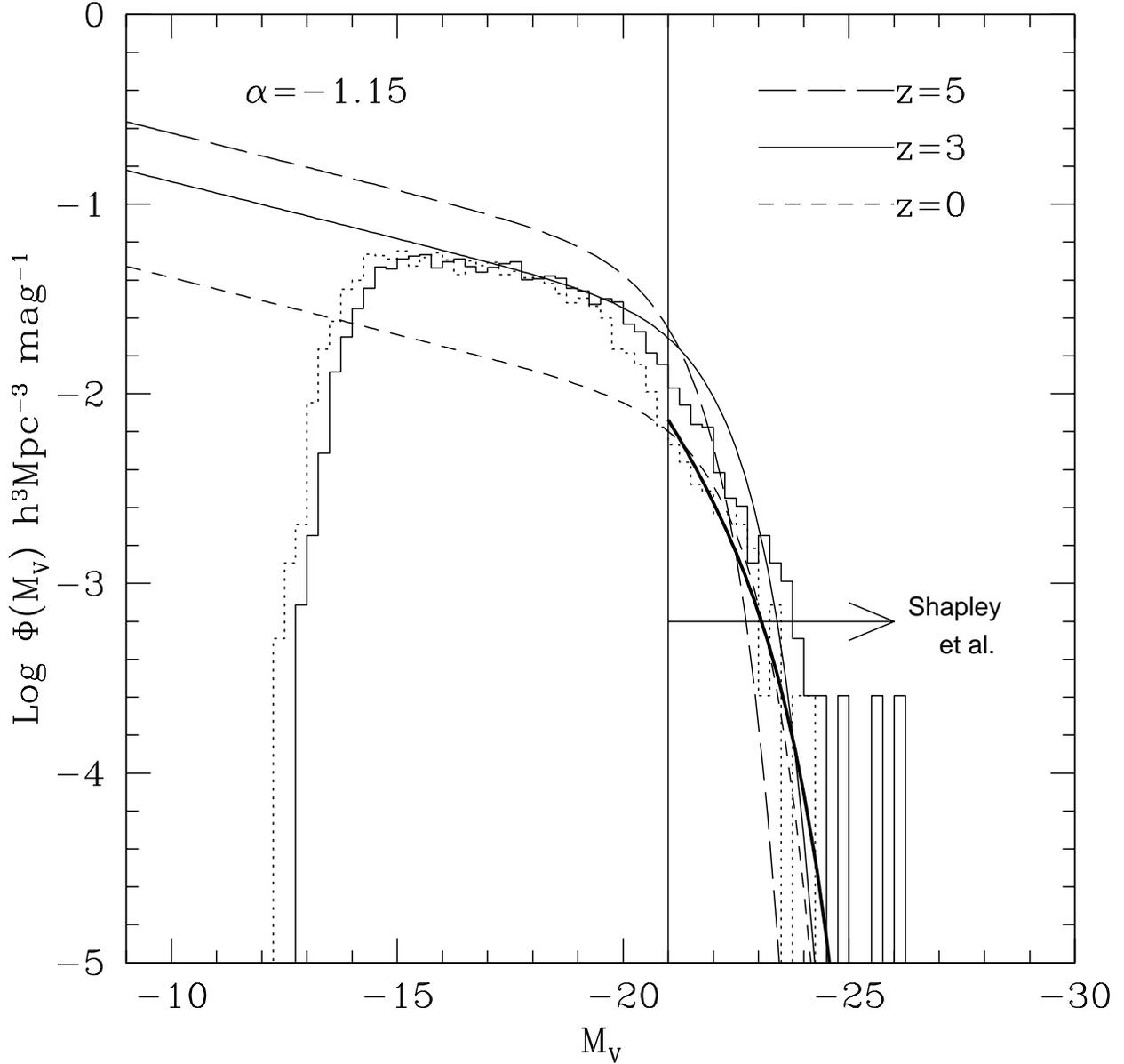}
\caption{
Solid histogram shows the rest-frame $V$ band luminosity function 
derived from the simulation at $z=3$ (before dust extinction). 
Short-dashed, solid, and long-dashed curves show the Schechter 
function fit at $z=0,3,$ and 5, respectively. 
See text for the fitting method.
The Schechter parameters are summarized in Table~\ref{table1}.
The heavy solid line is the observed luminosity function
by \citet{Shapley01}, and their detection limit ($M_V=-21.0$) 
in near-IR is shown by the vertical solid line and the arrow. 
The dotted histogram is the dust extincted luminosity function 
in the simulation. 
Current surveys sample only the bright-end of the luminosity 
function at $z=3$, and are in good agreement with the simulated one
with dust extinction.
\label{lf.eps}}
\end{figure}

In \Fig{lf.eps}, we show the computed rest-frame $V$ band luminosity 
function of galaxies at $z=3$ in the solid histogram
(before dust extinction).
The Schechter function fits at $z=0,3,$ and 5 are shown as the 
short-dashed, solid, and long-dashed curves, respectively.
The method for the Schechter fit is as follows: first we fix 
the faint-end slope to $\alpha=-1.15$ (This value is consistent with 
the empirical one in $B$ band at $z=0$. See Paper II.), then adjust 
the normalization $\Phi^*$ to fit the plateau at the faint end of 
the luminosity function, 
and then choose the characteristic magnitude $\Mstarv$ so that 
the integral of the Schechter function is equal to the total 
$V$ band luminosity density in the simulation box. 
This fitting method is adopted to take into account of the 
luminosity of a few overmerged objects in the simulation (see Paper II). 
At $z=3$, we obtain $\Phistar=2.70\times 10^{-2}h^3 \mpc^{-3}$
and $\Mstarv=-22.05$ ($h=0.67$). The Schechter parameters at other 
redshifts are summarized in Table~\ref{table1}. Similar table in 
the rest-frame $B$ band was reported in Paper II. 
The dotted histogram is the luminosity function after applying
the dust extinction model described in \S~\ref{section:selection_lbg}. 
When the same fitting procedure is applied to the dotted histogram, 
it gives $\Phistar=2.70\times 10^{-2}h^3 \mpc^{-3}$ and $\Mstarv=-21.28$ 
($h=0.67$). \citet{Shapley01} report the Schechter parameters of 
the rest-frame $V$ band luminosity function of LBGs to be:
$\alpha=-1.85\pm 0.15$, $\Mstarv=-22.21\pm 0.25 +5\log h$,
and $\Phistar=(0.18\pm 0.08)\times 10^{-2} h^3 \mpc^{-3}$,
which is shown as the thick solid line in \Fig{lf.eps}. 

\notetoeditor{Place \Fig{lf.eps} and Table~\ref{table1} here.}

At the bright end of the luminosity function, the 
agreement between the dust extincted computed luminosity function
(dotted) and the observed one (thick solid) is very good. 
This suggests that the intrinsic value of $\Phistar$ could be 
much larger than that given by \citet{Shapley01}.
The observational estimate of the faint end slope 
$\alpha=-1.85$ is significantly steeper than the simulated result, 
but this is perhaps due to the fact that they have only observed 
the bright end of the luminosity function.
Our result suggests that, with the magnitude limit of $M_V\sim -21$, 
one is not seeing the faint-end of the luminosity 
function sufficiently at $z=3$, and dust extinction exacerbates
the situation. Therefore it is observationally difficult to derive 
the Schechter parameters correctly with the current detection limit. 

A further detailed comparison of the parameters $\Mstarv$ and $\Phistar$
is meaningless because of the huge covariance between $\alpha$, 
$M^*$, and $\Phistar$. What counts here is the luminosity function 
itself, rather than the exact value of Schechter parameters. 
We consider that the agreement between the simulated and the observed
luminosity function is fair given the uncertainties involved
in the theoretical modeling of the simulation and observations. 
We remark that, in Paper II, we found a similar level of agreement 
in the rest-frame $B$ band at $z=0$ between the simulated luminosity 
function and the empirical estimates by the Sloan Digital 
Sky Survey \citep{Blanton00}  and the 2dF survey 
\citep{Cross01}. It is encouraging that we get this level of
agreement between the simulation and the observations without 
any fine-tuning of the parameters of the simulation. 

Our simulation predicts a rest-frame $V$ band comoving luminosity 
density of $\Ldenv=7.6\times 10^8 h \Lsunv\mpc^{-3}$ (before extinction)
and $3.8\times 10^8 h \Lsunv\mpc^{-3}$ (after extinction),
when the luminosity of all the galaxies in the simulation is summed up.
If the sum is taken only for the galaxies above the detection limit
of $M_V=-21.0$, we get $4.7\times 10^8h \Lsunv\mpc^{-3}$ (before extinction) 
and $1.7\times 10^8h \Lsunv\mpc^{-3}$ (after extinction).
\citet{Shapley01} give $\Ldenv=1.35\times 10^8 h \Lsunv\mpc^{-3}$ 
by integrating the observationally inferred luminosity function 
to the same detection limit.
The last two values above are consistent with each other
when uncertainties in both sides are considered.
Further detailed comparison of total luminosity density 
is hindered by the current detection limit.
We remark that the predicted $B$ band luminosity density 
and the stellar mass density at $z=0$ in our simulation are
consistent with the empirical estimates within the uncertainties 
(see Paper II).

\citet{Dave99} calculated the $R$ band luminosity function
by assuming a galactic extinction law with fixed $A_V=1.0$
for all galaxies in a SPH simulation, and argued that the 
result is in rough agreement with the observed luminosity 
function when dust extinction is taken into account, although
the simulated luminosity function was somewhat steeper.
A detailed comparison of Schechter parameters was not 
presented in their paper.


\section{Stellar Mass of LBGs}
\label{section:stellarmass_lbg}

\begin{figure}
\epsscale{1.0}
\plotone{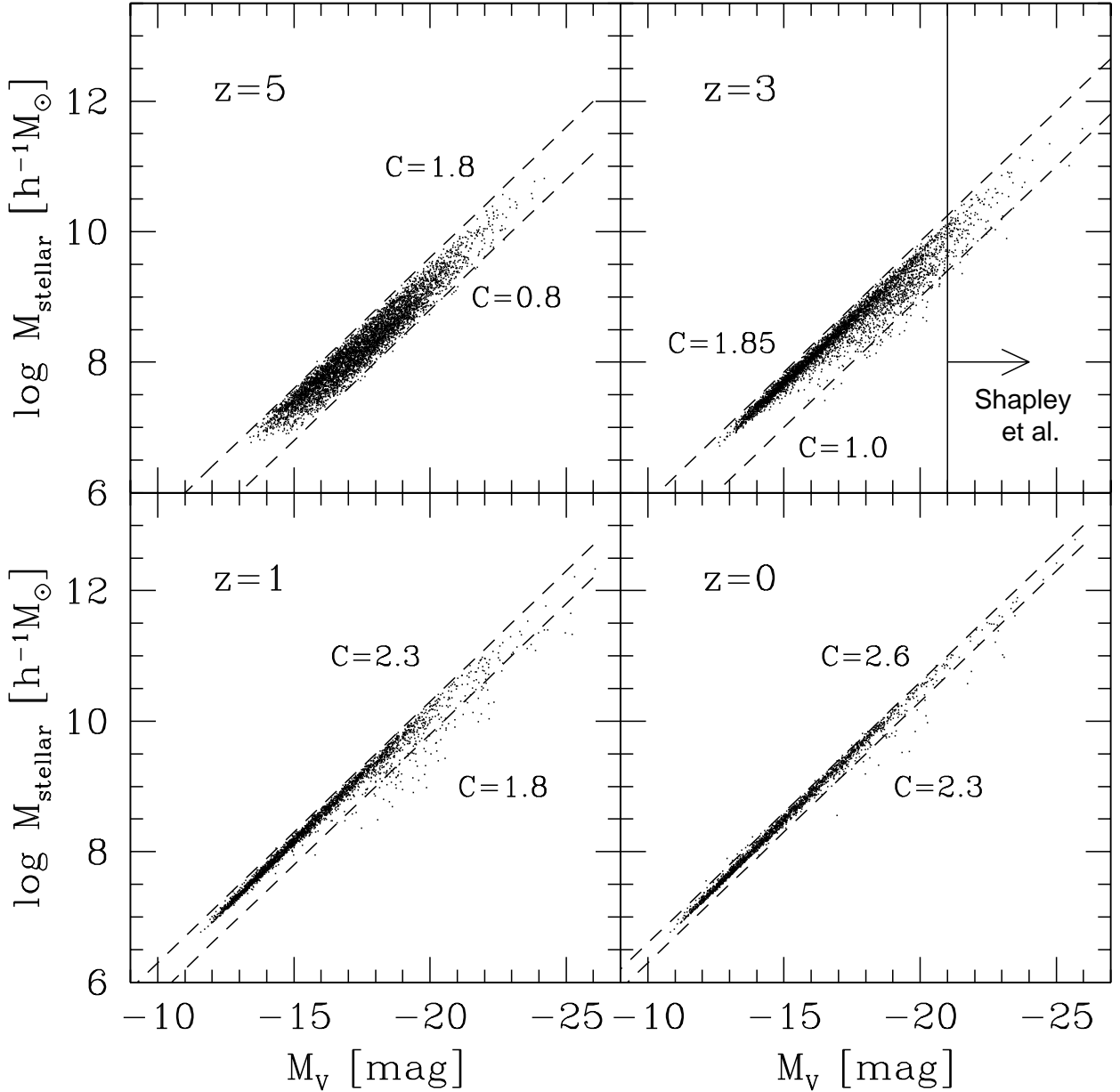}
\caption{
Stellar mass of galaxies vs. rest-frame $V$ band 
absolute magnitude (before dust extinction) at $z=5,3,1$, and 0. 
Each point in the figure represents one galaxy in the simulation.
The dashed lines are drawn by hand to roughly indicate the locus of the 
distribution, together with the values of the constant C, 
where $\log \Mstellar = -0.4 M_V + C$ and $\Mstellar$ is in units
of [$\hinv\Msun$].
The entire distribution shifts to the upper left
direction from high to low-redshift, as galaxies become 
more massive and less luminous. The wider distribution 
at high-redshift indicates more active star formation activity
and its variety. 
The detection limit of $M_V=-21.0$ \citep{Shapley01} is 
indicated by the vertical solid line and the arrow in $z=3$ panel.
The median stellar mass of all galaxies above this detection limit
is $10^{10}\himsun$.
\label{mass_VmagLBG.eps}}
\end{figure}

In our simulation, the luminosity and stellar mass of galaxies
correlate well, as shown explicitly in \Fig{mass_VmagLBG.eps}.
In each panel, the dashed lines are drawn by hand to roughly 
indicate the locus of the distribution, together with the 
value of the constant C, where $\log \Mstellar = -0.4 M_V + C$
and $\Mstellar$ is in units of $[\hinv\Msun]$.
The entire distribution shifts to the upper left
from high to low redshift, as galaxies become 
more massive and less luminous with time on average.
The scatter in the distribution is larger at higher redshift,
reflecting the younger age of stars and the variety in the star formation 
histories of high-redshift galaxies relative to low-redshift ones.

\notetoeditor{Place \Fig{mass_VmagLBG.eps} here.}

In the $z=3$ panel, the detection limit of $M_V=-21.0$ 
\citep{Shapley01} is indicated by the vertical solid line and 
the arrow. The intersection of the solid and the dashed lines 
indicates that the currently observed LBGs have stellar masses 
of at least $10^9\himsun$, and those at the detection limit 
are in the range $10^9-10^{10}\himsun$. 
The median stellar mass of all galaxies that are brighter than
$M_V=-21.0$ is $10^{10}\himsun$.
When dust extinction is taken into account, the distribution
becomes dimmer by about a magnitude, and the mass range
at the magnitude limit shifts upward slightly: $10^{9.5}-10^{10.5}\himsun$.
Our result is consistent with the derived stellar mass
by \citet{Shapley01}.


\section{Merger Trees and Star Formation Histories of LBGs at $z\geq 3$}
\label{section:mergertree_lbg}

\begin{figure}
\epsscale{1.0}
\plotone{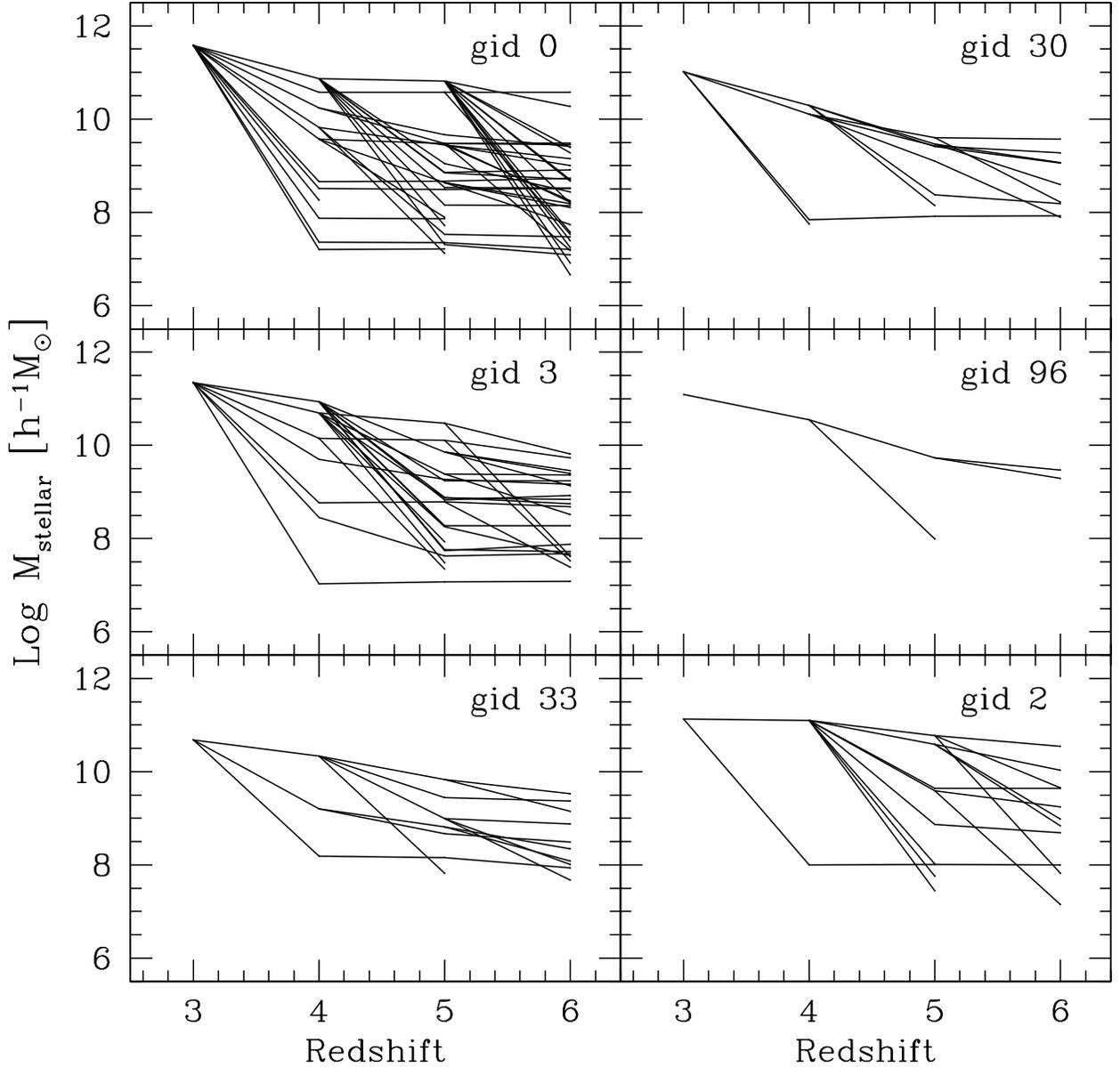}
\caption{
Merger trees of the brightest LBGs (intrinsic luminosity 
$M_V<-23.5$ at $z=3$) are shown
in the order of their rest-frame $V$ band luminosity at $z=3$ 
from top left to bottom right. The ordinate is the stellar mass of 
galaxies. Each leg in the figure connects two objects before and 
after a merger event that took place in between the two redshifts.
The corresponding composite star formation histories are shown in 
\Fig{SFhist_LBG.eps}. See text for the discussion on each galaxy. 
\label{z3treeplot.eps}}
\end{figure}

In \Fig{z3treeplot.eps}, we show the merger trees of some of the 
brightest LBGs at $z=3$ in the simulation in the order of 
their rest-frame $V$ band luminosity at $z=3$
(galaxy ID [gid]$=0, 30, 3, 96, 33, 2$). 
All of these galaxies are also bright in the rest-frame far-UV,
and should be visible in the current far-UV selected galaxy 
observation as well. 
The figure shows the stellar mass of galaxies at $z=3,4,$ and 5
in the ordinate. The merger events are shown by connecting 
the points at each epoch.
Note that the actual merger events might have taken place 
anytime between the two redshifts. 

\begin{figure}
\epsscale{1.0}
\plotone{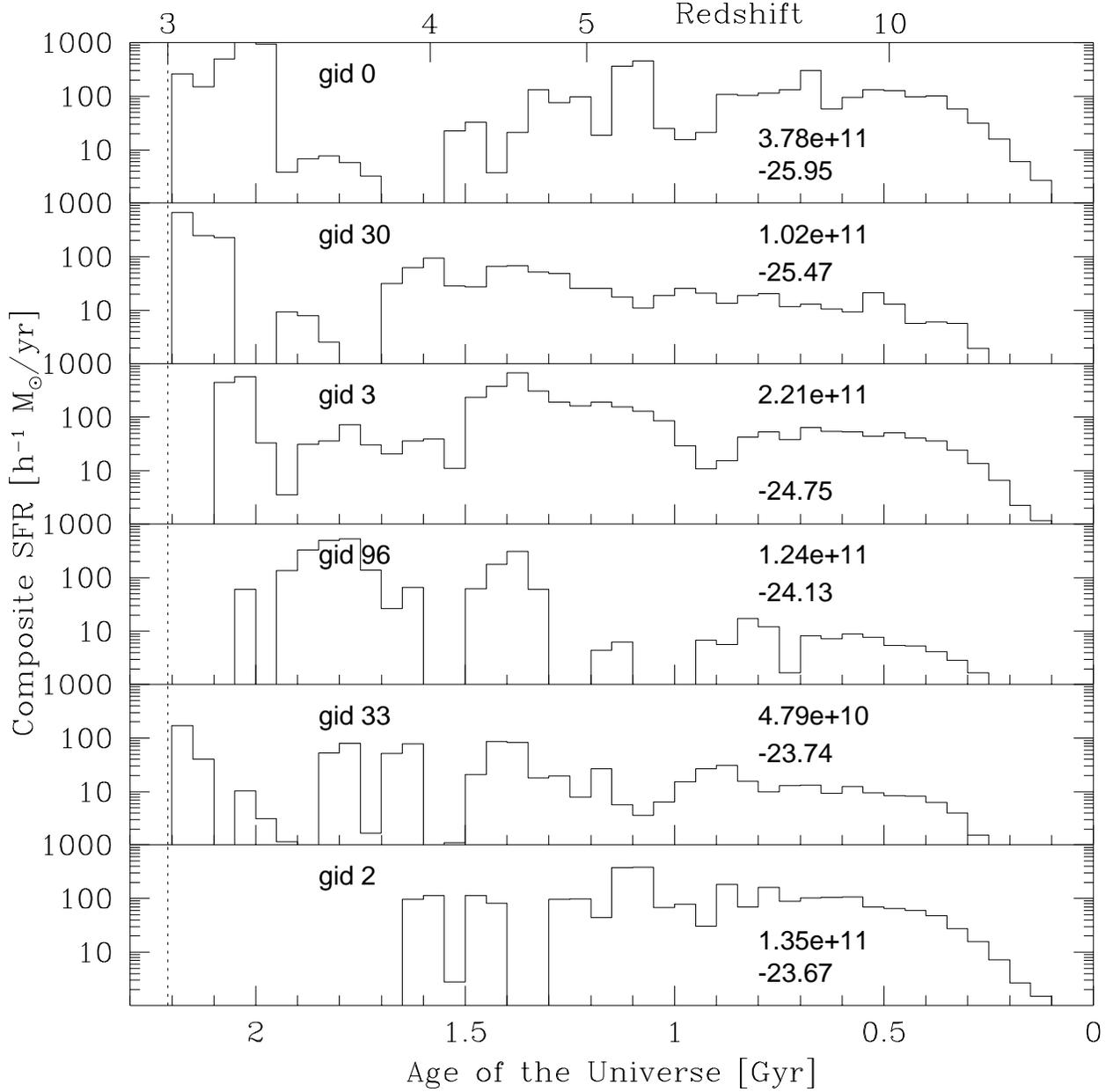}
\caption{
Composite star formation histories of LBGs in \Fig{z3treeplot.eps}
in the order of the rest-frame $V$ band luminosity at $z=3$
from top to bottom. The bin size of the histogram is 50 Myr. 
Note that this is the composite star formation rate of all 
progenitors which fall into a given LBG at $z=3$.
The values of the stellar mass (in units of $\Msun$) and the 
rest-frame $V$ band absolute magnitude of the object at $z=3$ 
are given in each panel. The vertical dotted line on the left 
indicates $z=3$. See text for the discussion for each galaxy. 
\label{SFhist_LBG.eps}}
\end{figure}

In \Fig{SFhist_LBG.eps}, we show the composite star formation 
histories of the progenitors which end up in these LBGs,
in the decreasing order of the rest-frame $V$ band luminosity 
from top to bottom. In each panel, the values of stellar mass 
(in units of $\Msun$) and rest-frame $V$ band absolute magnitude 
of each object at $z=3$ are indicated. The bin size of the 
histogram is 50 Myr.

\notetoeditor{Place \Fig{z3treeplot.eps} and \Fig{SFhist_LBG.eps} here.}

Now let us describe several selected objects more closely. 

The galaxy gid=0 is the most massive and the brightest object in 
the simulation at $z=3$, and has $\Mstellar=3.78\times 10^{11}\himsun$.
It might be suffering from the overmerging problem at $z=3$ 
(i.e. could be a few separate objects in reality). 
It has experienced multiple merger events from $z=5$ to $z=3$.
Its star formation history (top panel of \Fig{SFhist_LBG.eps}) tells 
us that the progenitors of this object have continuously formed 
stars from $z=10$ to $z=4$.
It has experienced a significant starburst in between 
$z=4$ and $z=3$ which lasted over a period of 100~Myr.
This galaxy grows into a group of galaxies with 
$\Mstellar=2.6\times 10^{12} \himsun$ at $z=0$, as we will see in 
the next section. 

The galaxies gid=30, 3, and 96 all merge into a same cluster 
at $z=0$ with $\Mstellar=6.0\times 10^{12}\himsun$, as we will show 
in the next section.  
Interestingly, gid=30 has the smallest stellar mass among the three, 
but it is the brightest because it had a starburst closer to $z=3$ 
than the other two.

The galaxy gid=96 offers an interesting example, where we see a major 
starburst at $z=3-4$ but no corresponding major merger event during the 
same period. A significant gas infall without a galaxy merger event must
have taken place in this case. This galaxy only experiences two merger 
events in the interval $z=4-5$ and $z=5-6$. 

The galaxy gid=33 is interesting because it grows into 
a galaxy of $\Mstellar =1.9\times 10^{11}\himsun$ at $z=0$, 
a Milky Way sized galaxy with the color of $B-V=0.82$,
typical of a S0a galaxy \citep{Roberts94}. 
It experiences a moderate number of merger events at $z>3$, 
and has continuously formed stars over a period of 1 Gyr.

The galaxy gid=2 is also interesting in that it has almost 
no star formation activity in the interval $z=3-4$, but is 
still bright enough at $z=3$ because of the significant star 
formation at $z>4$ associated with multiple merger events.
This object merges into the same group at $z=0$ as gid=0 does.

To summarize, the majority of LBGs that are intrinsically brighter 
than $M_V=-23$ at $z=3$ have experienced intensive star formation 
activities ($\geq 100\Msun /\yr$) within 500~Myr prior to $z=3$, 
often lasting for a period of $\sim 100$~Myr, resulting in a stellar 
mass of $>10^{10}\Msun$. Besides the starbursts, the progenitors 
of these bright LBGs have continuously formed 
stars over a period of 1~Gyr since $z=10$ when they are coadded.
Qualitatively, these histories seem to be consistent with the star 
formation histories inferred from recent near-IR observations 
of LBGs \citep{Papovich01, Pettini01, Shapley01}.
Most of the starbursts in the simulation seem to be associated with 
merger events, but this is not necessarily so, as in the case of gid=96. 
Milder gas infall not associated with a merger event has taken 
place in such cases.
At $z=5$, the stellar mass of the progenitors of LBGs ($M_V<-21.0$ at $z=3$) 
ranges from $10^{6.9}\himsun$ to $10^{10.8}\himsun$, with the median 
value of $10^{8.6}\himsun$.


\section{The Fate of LBGs at $z\leq 3$}
\label{section:fate_lbg}

\subsection{Merger Tree and Star Formation History}

\begin{figure}
\epsscale{1.0}
\plotone{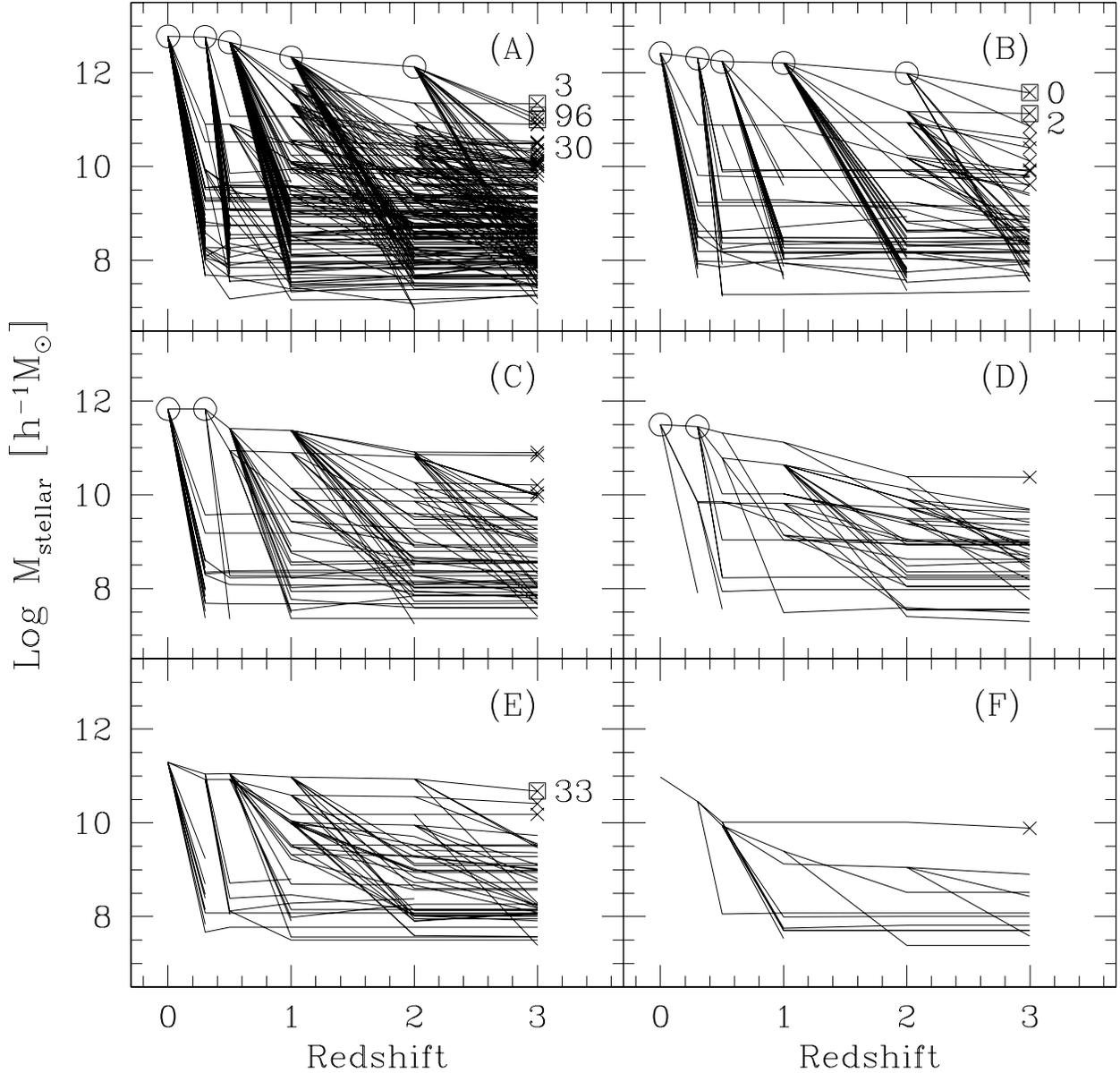}
\caption{
Merger trees of the present day cluster, groups, and $L^*$ galaxies
are shown in the order of their stellar mass at $z=0$ from top 
to bottom. The ordinate is the stellar mass of galaxies. 
The LBGs that are brighter than $M_V=-21.0$ are indicated by 
the crosses at $z=3$. The open circles in panels (A)$-$(D) 
indicate that the objects are suffering from the overmerging problem 
in the simulation, and represent clusters/groups of galaxies as a whole.
The corresponding star formation histories are shown in 
\Fig{SFhist_z0.eps}. See text for the discussion on each galaxy. 
\label{z0treeplot.eps}}
\end{figure}

In \Fig{z0treeplot.eps}, we show the merger histories of representative 
objects which end up as clusters (panels A \& B), groups (panels C \& D), 
and $L^*$ galaxies (panels E \& F) at $z=0$.
The LBGs that are brighter than $M_V=-21.0$ are indicated by the  
crosses at $z=3$. Some of the brightest LBGs that were described 
in \Fig{z3treeplot.eps} are indicated by the open squares at $z=3$,
with their galaxy IDs on the side. The open circles in panels 
(A)$-$(D) indicate that the objects are 
suffering from the overmerging problem in the simulation, 
and represent clusters/groups of galaxies as a whole.

\notetoeditor{Place \Fig{z0treeplot.eps} here.}

The panels (A) and (B) show that a present day cluster
of galaxies contains roughly $\gtsim 10$ LBGs as their progenitors,
as indicated by the crosses at $z=3$ in the figure.
They continue to collect numerous less massive objects with 
$\Mstellar=10^7 - 10^9\Msun$ at $z=0-3$.
We find that the majority of LBGs that are intrinsically 
brighter than $M_V=-23$ at $z=3$ merge into clusters/groups of galaxies 
at $z=0$ in our simulation. 

The panels (C) and (D) show that a present day group of 
galaxies typically contains one to $\ltsim 10$ LBGs as their
progenitors. We find that roughly one half of LBGs with 
$-23\ltsim M_V \ltsim -22$ merge into groups/clusters of galaxies, 
while the other half evolve into typical $L^*$ galaxies with 
$\Mstellar\sim 10^{11}\Msun$ as shown in the panels (E) and (F). 
Present day galaxies with $\Mstellar\gtsim 10^{11}\Msun$
typically have one to a few LBGs as their precursors which are brighter 
than $M_V=-21.0$ and have $\Mstellar\gtsim 10^{10}\Msun$ at $z=3$.

\begin{figure}
\epsscale{1.0}
\plotone{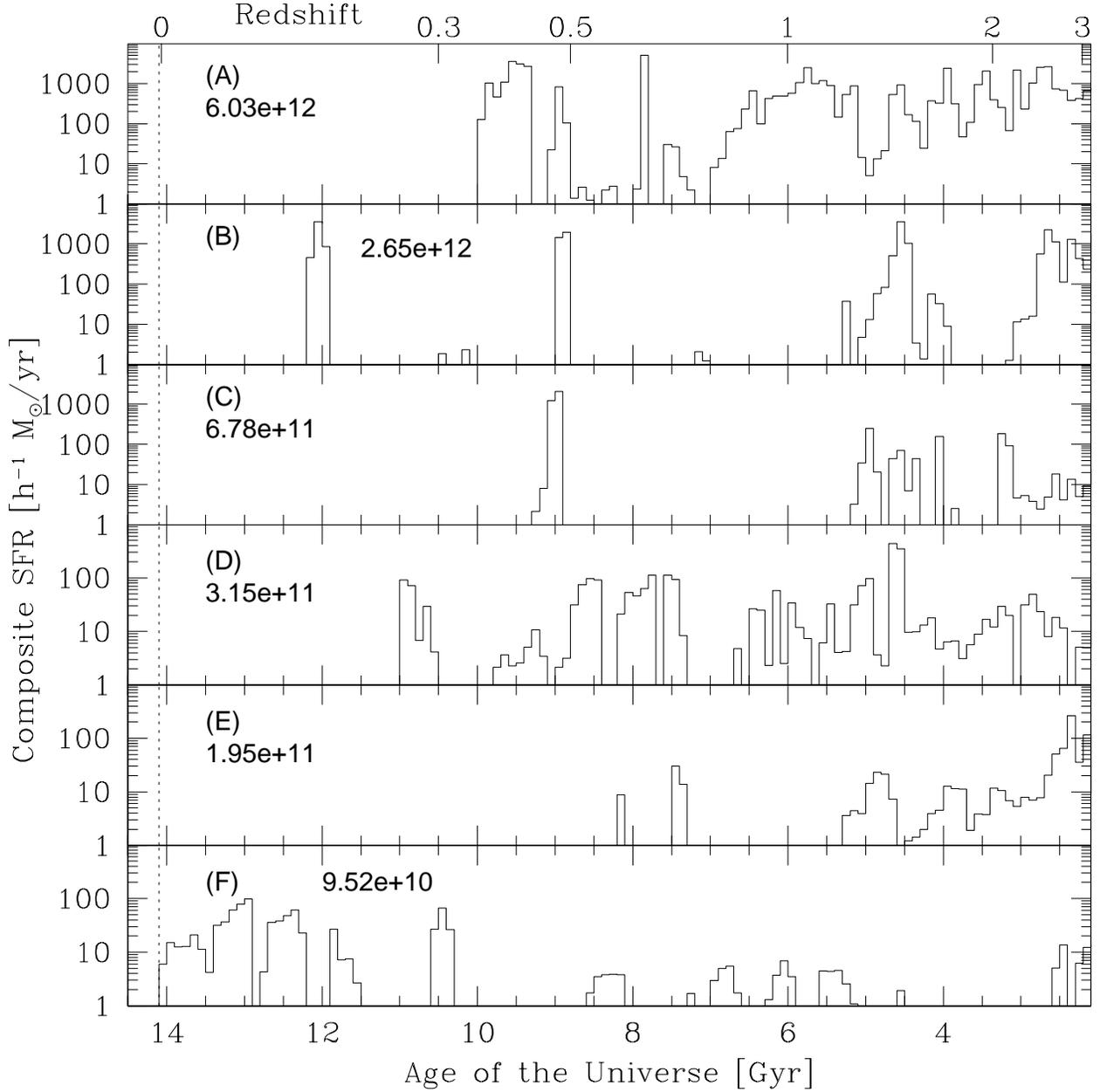}
\caption{
Composite star formation histories of objects shown in 
\Fig{z0treeplot.eps} as a function of the age of the universe.
The bin size of the histogram is 100 Myr. 
Note that the histograms are the composite star formation rate
of all progenitors that fall into the same object at $z=0$.
The value of the stellar mass of each object at $z=0$ is shown 
in each panel in units of $\hinv\Msun$. The vertical dotted line 
indicates the present epoch in the simulation. See text for the 
discussion on each galaxy. 
\label{SFhist_z0.eps}}
\end{figure}

The corresponding composite star formation histories for 
all cases are presented in \Fig{SFhist_z0.eps} as functions of 
the age of the universe and redshift.
The exceptionally high star formation rates
seen for (A), (B), and (C) are because these histograms
are the composite of all progenitors which later fall 
into the same cluster/group. The stellar mass of the object at $z=0$ 
is indicated in each panel.

One sees that the star formation activity is more episodic
at lower redshifts. In very massive systems (panels A$-$C),
we see no quiescent star formation at low-redshift, 
which agrees with the observed absence of 
star formation in clusters of galaxies at late times.
This is presumably due to the inefficient cooling of the gas 
in massive systems because of the high temperature of the
shock-heated gas \citep{Rees77, Silk77, Blanton99}.

\notetoeditor{Place \Fig{SFhist_z0.eps} here.}

Panel (D) has relatively frequent star formation 
activity up to $z\sim 0.2$. It only contains a single LBG
which is brighter than $M_V=-21.0$ at $z=3$. The total 
stellar mass of this system at $z=0$ is $3.15\times 10^{11}\himsun$, 
with a color of $B-V=0.86$, typical of a S0 galaxy \citep{Roberts94}.

Panel (E) offers an interesting example of a system with 
stellar mass of $\sim 2\times 10^{11}\himsun$ at $z=0$, 
with a color of $B-V=0.82$, typical of a S0a galaxy \citep{Roberts94}. 
Despite of the multiple accretion of small systems up to 
the present epoch, it hardly forms stars at $z<1$. 
It contains a few LBGs ($M_V<-21.0$) as its precursor at $z=3$.

A counter-example to (E) is shown in panel (F), where 
more star formation activity is seen with less merger events
when compared with (E). This system has only one LBG as its
precursor at $z=3$, and this LBG remains untouched until $z=1$.
Major merger events take place at $z<0.5$, inducing
significant star formation, and the stellar mass of the 
system increases dramatically at $z<0.5$.

\subsection{Spheroids as the Descendants of LBGs}

\begin{figure}
\epsscale{1.0}
\plotone{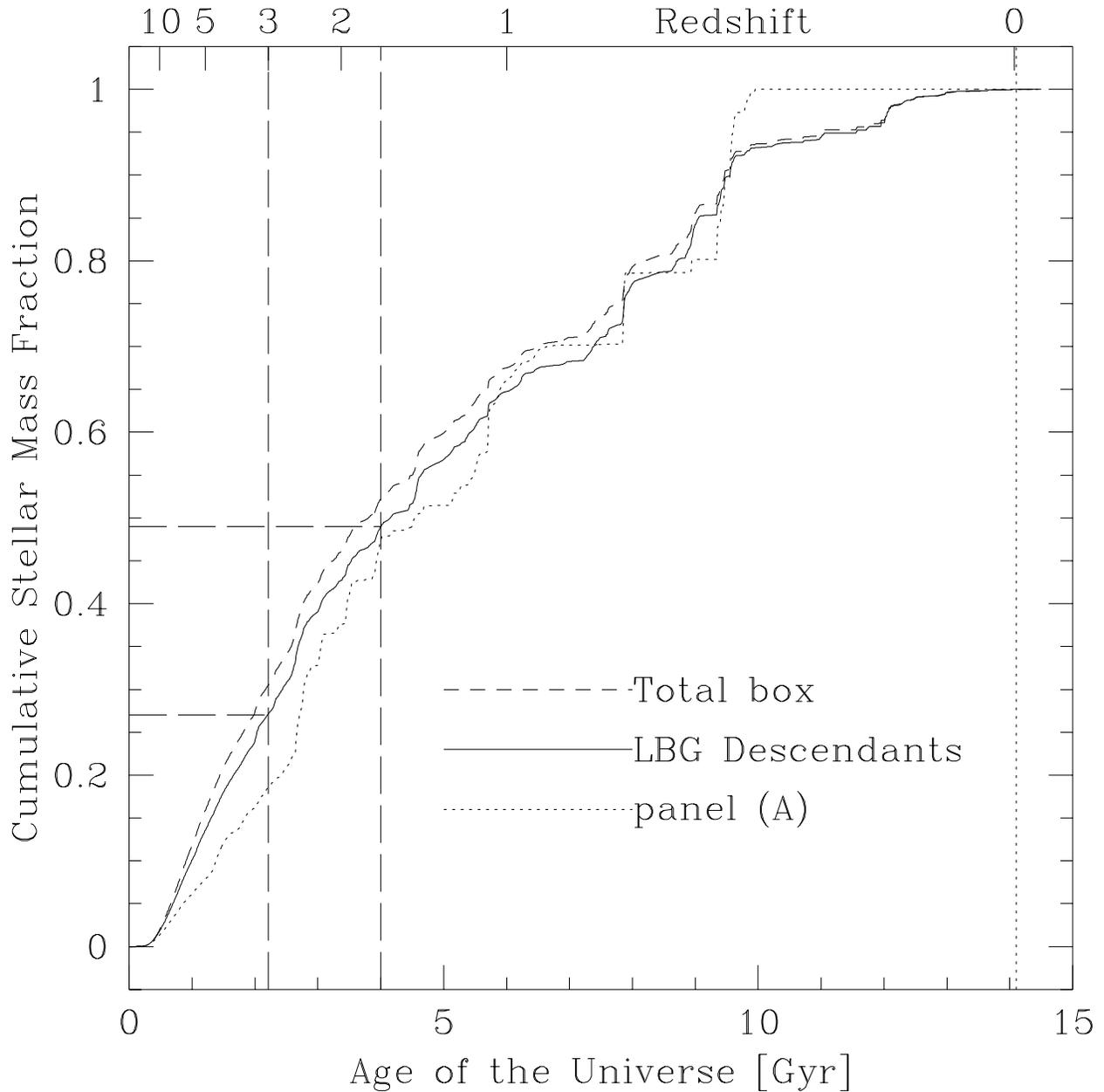}
\caption{
Cumulative stellar mass fraction as a function of 
the age of the universe and redshift for the total box (short dashed), 
descendants of the LBGs which are brighter than $M_V=-21.0$ at $z=3$
(solid), and the cluster shown in panel (A) of \Fig{z0treeplot.eps}.
The long dashed lines indicate that the present day descendants of 
LBGs have formed $\sim 30$\% of their stellar mass 
by $z=3$, and $\sim 50$\% of their present day stellar population is
10 Gyr old, in favor of the scenario that LBGs are the precursors
of the present day spheroids. 
\label{lbg_starcum.eps}}
\end{figure}

In \Fig{lbg_starcum.eps}, we show the cumulative stellar mass fraction
as a function of the age of the universe for the total box (short dashed), 
descendants of the LBGs which are brighter than $M_V=-21.0$ at $z=3$
(solid), and the cluster shown in panel (A) of \Fig{z0treeplot.eps}.
The long dashed lines indicate two epochs: $z=3$ and $t=4$ Gyr,
where $t$ is the age of the universe.
The intersections with the solid curve show that the 
descendants of LBGs today have formed $\sim 30$\% of their stellar mass 
by $z=3$, and $\sim 50$\% by $t=4$ Gyr. In other words, 
50\% of the present day stellar population in LBG descendants 
is 10 Gyr old. Together with the merger trees presented in the
previous subsection, this suggests that the stars we see as 
LBGs at $z=3$ will evolve into massive old stellar component 
of the present day galaxies; i.e. the `spheroids' 
(bulges of spirals and luminous ellipticals).
The observed compactness of LBGs \citep[$\ltsim 4\kpc$;][]{Lowenthal97,
Giavalisco96} is consistent with this idea. 
Many authors also prefer the spheroid scenario to the 
low-mass starburst scenario \citep{Lowenthal97} and the satellite
starburst \citep{Somerville01} from other considerations,
such as a chemodynamical model which combines chemical evolution
model with 1 dimensional hydrodynamics \citep{Friaca99},
or a semianalytic model which predicts sizes, kinematics, and 
star formation rates of LBGs \citep*{Mo99}.

Unfortunately, due to limited spatial resolution of our simulation, 
we cannot address the question of what happens to the LBGs that fall
into clusters. \citet{Governato01} have explored this issue
by combining a semianalytic galaxy formation model with a 
high-resolution N-body simulation of a cluster. They find that the 
descendants of LBGs in clusters can be identified with a central 
cD galaxy and giant elliptical galaxies within the central 60\% of the 
virial radius of a cluster.


\section{Star Formation Rate of Galaxies: Instantaneous vs. Continuous}
\label{section:sfr_lbg}

\begin{figure}
\epsscale{1.0}
\plotone{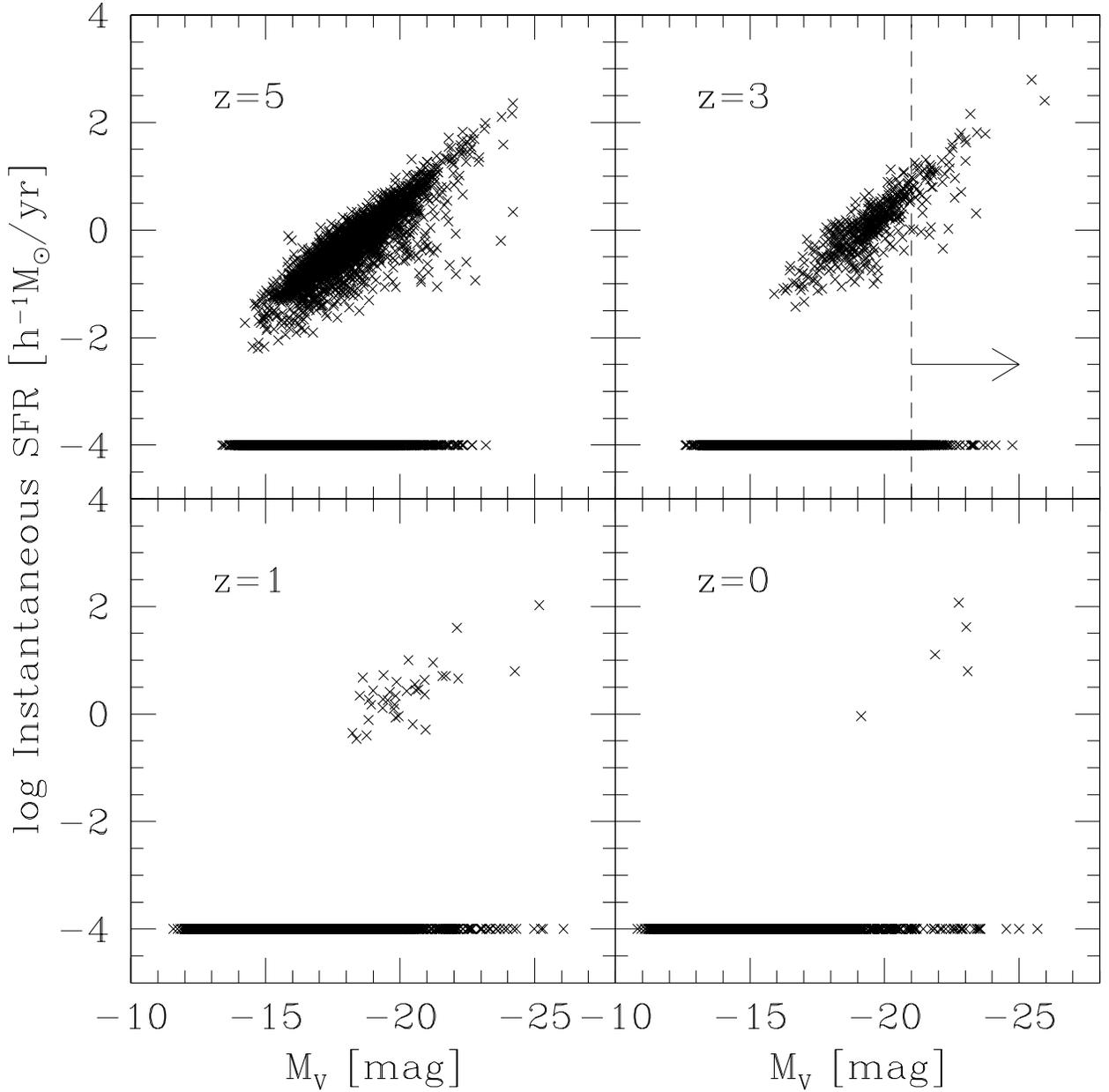}
\caption{Instantaneous star formation rate of galaxies 
at $z=0, 1, 3$, and 5 as a function of rest-frame $V$ band 
absolute magnitude (before dust extinction), calculated by 
averaging over 20~Myr at each epoch. 
Those which are assigned $10^{-4} [\himsun/\yr]$
happen to have no star formation at that epoch, due to the 
intermittent nature of the star formation in the simulation.
The vertical dashed line and the arrow at $z=3$ indicates 
the detection limit of $M_V=-21.0$ \citep{Shapley01}.
\label{rate_LBGmag.eps}}
\end{figure}

In \Fig{rate_LBGmag.eps}, we show the instantaneous 
star formation rate of galaxies at $z=0,1,3,$ and 5,
calculated by averaging over 20~Myr at each epoch,
as a function of rest-frame $V$ band absolute magnitude
(before dust extinction).
We have confirmed that the result is robust against the 
change in the time-scale of the averaging. 
Those which are assigned ${\rm SFR}=10^{-4} [\himsun/\yr]$
happen to have no star formation at that particular epoch,
due to the intermittent nature of the star formation in 
the simulation.
Note that ongoing star formation due to recycled matter
within a galaxy might not be captured in this simulation
due both to limited spatial resolution and the nature of
the star formation recipe implemented in the simulation. 
The vertical dashed line at $z=3$ indicates the detection 
limit of $M_V=-21.0$ \citep{Shapley01}.

\notetoeditor{Place \Fig{rate_LBGmag.eps} here.}

There is a relatively strong correlation between the 
instantaneous SFR and the luminosity of galaxies at $z\gtsim3$: 
more luminous galaxies are forming stars more actively.
Because of the correlation between luminosity and stellar mass
shown in \Fig{mass_VmagLBG.eps}, this means that more massive
galaxies are forming stars more actively on average in our simulation.
Galaxies that are brighter than $M_V=-21.0$ take a wide range of 
instantaneous SFR; from $1\Msun/\yr$ to a few $100 \Msun/\yr$.  
Galaxies at the detection limit have values ranging from 
$1\Msun/\yr$ to $\sim 30\Msun/\yr$.  
At lower redshifts ($z\leq1$), star formation becomes
inactive on average, as was shown in Paper~I. 

\begin{figure}
\epsscale{1.0}
\plotone{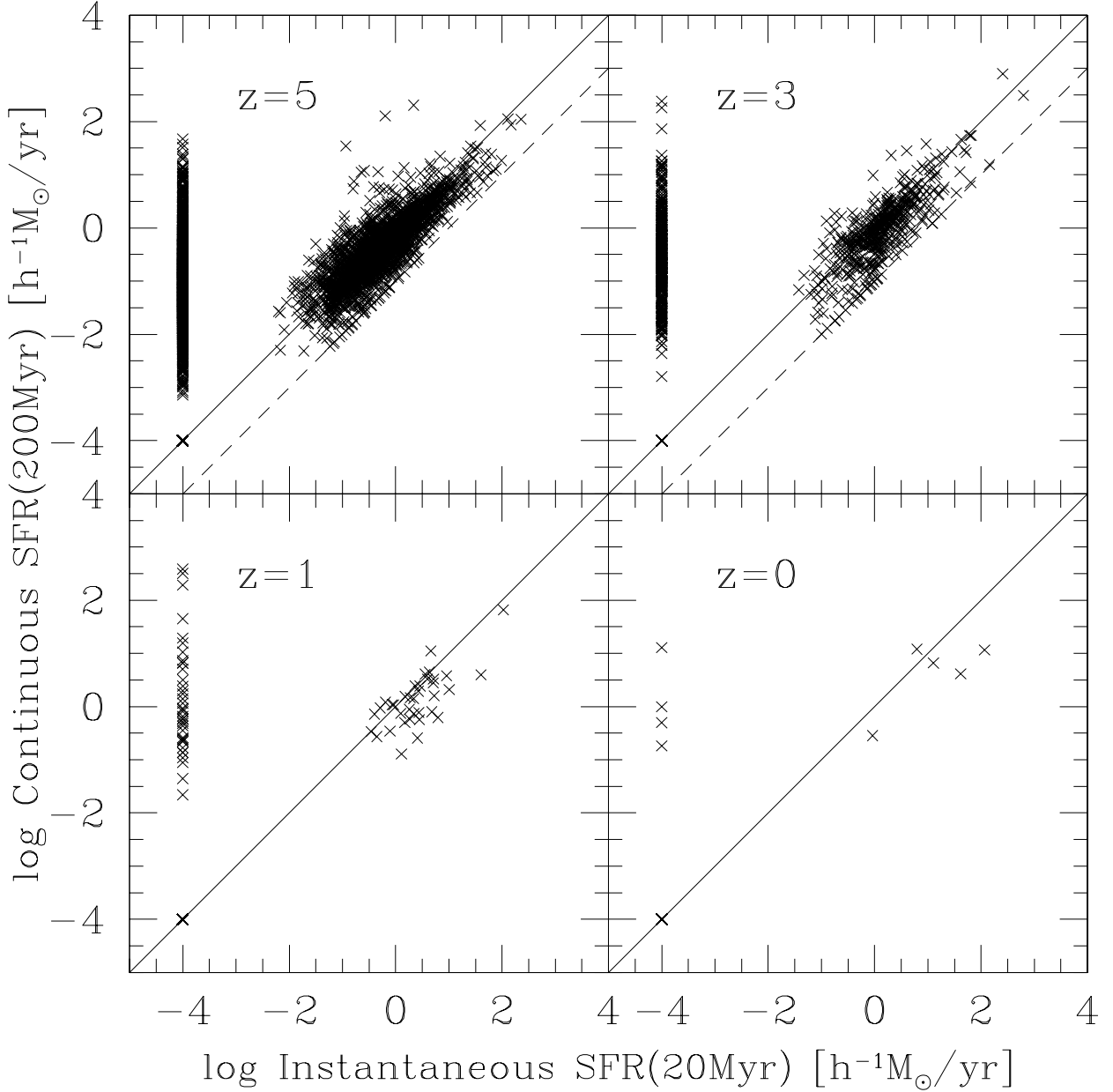}
\caption{`Instantaneous SFR' (averaged over 20~Myr) vs. 
`Continuous SFR' (averaged over 200~Myr) of galaxies 
in the simulation.  The diagonal solid line
indicates where the two SFRs take the same value; i.e. the 
star formation is continuous over at least 200~Myr
at the same level of SFR(20Myr).
The dashed lines in $z=5$ and $z=3$ panels indicate
the relation ${\rm SFR(200Myr)}=0.1\times {\rm SFR(20Myr)}$.
Bright LBGs occupy the upper region of the distribution. 
Some LBGs have episodic star formation, while others 
form stars continuously. 
The galaxies which are assigned $10^{-4} [\himsun/\yr]$
happen to have no star formation at that epoch, due to the 
intermittent nature of the star formation in the simulation.
\label{sf_sf.eps}}
\end{figure}

In \Fig{sf_sf.eps}, we plot `Instantaneous SFR(20Myr)' (averaged 
over 20~Myr) vs. `Continuous SFR(200Myr)' (averaged over 200~Myr)
of galaxies in the simulation. The diagonal solid line
indicates where the two SFRs take the same value; i.e. the 
star formation is continuous over at least 200~Myr
at the same level of SFR(20Myr).
The dashed lines in $z=5$ and $z=3$ panels indicate
the relation ${\rm SFR(200Myr)}=0.1\times {\rm SFR(20Myr)}$. 
At $z=5$ and 3, the lower envelope of the distribution 
follows the dashed line, indicating that in some galaxies
star formation is episodic; i.e. the level of SFR(20Myr) 
is not maintained over the period of 200~Myr. 
Therefore, in our simulation, star formation is episodic 
in some bright LBGs, while in others it is continuous.
There is no clear-cut distinction between the two: 
both episodic and continuous star formation take place.
The galaxies that are brighter than $M_V=-21.0$ at $z=3$
occupy the upper end of the distribution.

\notetoeditor{Place \Fig{sf_sf.eps} here.}


\section{Metallicity of LBGs}
\label{section:metal_lbg}

\begin{figure}
\epsscale{1.0}
\plotone{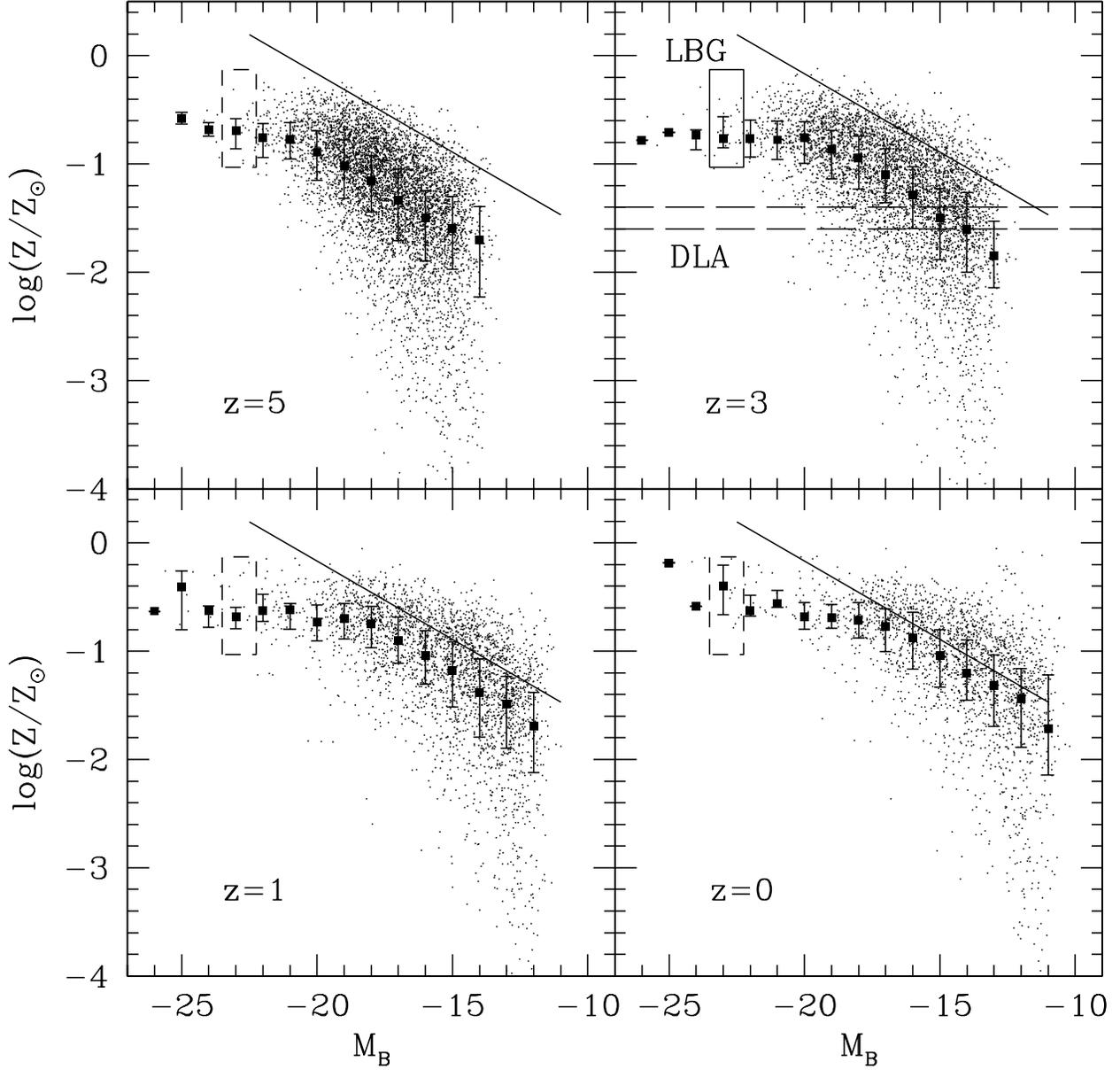}
\caption{
Mean stellar metallicity of galaxies vs. rest-frame $B$ band absolute 
magnitude (intrinsic) at redshifts $z=5,3,1$, and 0. The solid square 
points are the median in each magnitude bin, and the error bars are 
the quartiles. The solid box at $z=3$ indicates the range of metallicity 
of the LBGs given by \citet{Pettini01}. The same box is indicated by 
the dashed line at other epochs. The long-dashed horizontal lines at 
$z=3$ indicate the typical values of metallicity taken by the damped 
Lyman $\alpha$ systems \citep[$\sim 1/30 \Zsun$;][]{Pettini01}.
The slanted solid line is the best-fit to the observational data 
compiled by \citet{Kobulnicky99} for galaxies at $z=0-0.5$. The entire 
distribution shifts to lower metallicity at higher redshifts.
\label{metal_LBG.eps}}
\end{figure}

In \Fig{metal_LBG.eps}, we show the mean stellar metallicity 
of galaxies in the simulation vs. intrinsic rest-frame $B$ 
band absolute magnitude at redshifts $z=0,1,3,$ and 5.
The result at $z=0$ was presented in Figure~6 of Paper~I
with galaxy stellar mass as the x-axis.
The solid square points are the median in each magnitude bin, 
and the error bars are the quartiles. 
The slanted solid line is the best-fit to observational data 
compiled by \citet{Kobulnicky99} for galaxies at $z=0 - 0.5$, 
which shows good agreement with our result at $z=0$ in the 
range $-17\ltsim M_B\ltsim -10$. 
The distribution at $z=0$ deviates from the best-fit
line at $M_B\ltsim -17$, where we do not have enough statistics 
in the simulation due to limited box size. 
However, the flattening of the distribution at bright magnitudes 
at $z=3$ is likely to be real. 
The entire distribution shifts to lower metallicity 
at higher redshifts. This can be understood from Figure~5 
of Paper~I, where we showed that the stellar metallicity 
distribution widens significantly at $z\gtsim 3$, taking the values 
ranging from $10^{-4}\Zsun$ to $3.0\Zsun$. The physical reason for 
this effect is that the metallicity is a strong function of 
local overdensity \citep{Gnedin98, CO99}. 
As time progresses, the universe becomes more clumpy, 
and star formation takes place only in moderate to high overdensity 
regions where gas is already polluted by metals, resulting 
in a narrower scatter of metallicity at late times.

The solid box at $z=3$ indicates the range of metallicity 
inferred from the oxygen abundance of LBGs \citep{Pettini00}.
The same box is indicated by the dashed line at other redshifts.
Our model naturally explains the sub-solar metallicity of LBGs;
they take metallicities of $0.1-1.0\Zsun$ because they are 
the most luminous galaxies at $z=3$.

The long-dashed horizontal lines at $z=3$ indicate the 
typical values of metallicity taken by damped Lyman $\alpha$
systems (DLAs; $Z\approx (1/30)\Zsun$, \citet{Pettini01}). 
LBGs typically have higher metallicity than DLAs.
A semianalytic study by \citet{Mo99} suggests that DLAs are extended,
high angular momentum, low surface brightness, rotationally 
supported disks, while LBGs are low angular momentum, compact,
high surface brightness, possibly spheroid systems with high
star formation rate. High star formation rate
in compact LBGs results in a radial metallicity gradient
in a galaxy, with the gas in the outer region having much 
lower metallicity \citep{Shu00}.
The connection between LBGs and DLAs in our simulation 
will be discussed in detail in a separate paper \citep{Cen01}.

\notetoeditor{Place \Fig{metal_LBG.eps} here.}


\section{Conclusions}
\label{section:conclusion_lbg}

Using a large-scale hydrodynamic simulation,
we study the merger tree, the star formation history, 
the stellar content, and the metallicity of LBGs.
We focus our attention on galaxies that are intrinsically
brighter than $M_V=-21.0$ at $z=3$, and investigate the broad 
range of properties of such galaxies in accordance with recent
near-IR observations \citep{Papovich01, Pettini01, Shapley01, Rudnick01}.

The computed rest-frame $V$ band luminosity function of galaxies at 
$z=3$ is presented, and a fair agreement with the observed one is 
found, given the uncertainties in the simulation, the observations,
and the theoretical modeling including the initial mass function
and the population synthesis model. 
We find that the observed luminosity function by 
\citet{Shapley01} agrees well with the dust extincted simulated
luminosity function, and that the intrinsic $\Phistar$ could
be much larger than the value given by \citet{Shapley01}.

The predicted comoving luminosity density in the rest-frame $V$ 
band at $z=3$ is larger than the observed value of \citet{Shapley01} 
when integrated down to the current detection limit, but is
consistent within the uncertainties when dust extinction is taken 
into account. A detailed comparison of the total luminosity density 
at $z=3$ is hindered by the current detection limit, as the Schechter 
parameters cannot be determined reliably. 

The stellar mass of LBGs that have $M_V\simeq -21.0$ at $z=3$ 
lie between $10^{9.5}\himsun$ and $10^{10.5}\himsun$, 
when dust extinction is taken into account. The median stellar mass
of all LBGs that are intrinsically brighter than $M_V=-21.0$
at $z=3$ is $10^{10}\himsun$, in good agreement with the value 
inferred from a near-IR observation \citep{Shapley01}.
At $z=5$, the stellar mass of the progenitors of LBGs 
($M_V<-21.0$ at $z=3$) ranges from $10^{6.9}\himsun$ to 
$10^{10.8}\himsun$, with a median of $10^{8.6}\himsun$.

We find that the majority of LBGs brighter than $M_V=-23$
at $z=3$ fall into groups and clusters of galaxies by $z=0$.
Roughly one half of LBGs with $-23\ltsim M_V\ltsim -22$ merge into 
groups/clusters of galaxies, while the other half
evolve into typical $L^*$ galaxies with $\Mstellar\approx 10^{11}\Msun$. 
Present day galaxies with $\Mstellar\gtsim 10^{11} \Msun$
typically have a few LBGs as their progenitors 
brighter than $M_V=-21.0$, with $\Mstellar\gtsim 10^{10} \Msun$.
Descendants of LBGs at the present epoch formed
$\approx 30$\% of their stellar mass by $z=3$, and $\approx 50$\% 
of their present stellar population is 10~Gyr old. 
The existence of a significant old stellar population suggests that 
LBGs may be the precursors of the present day spheroids. 

Our simulation suggests that LBGs that are brighter than $M_V=-23$
at $z=3$ have experienced a burst of star formation 
($\geq 100\Msun /\yr$) within 500~Myr prior to $z=3$, 
often lasting for a period of $\approx 100$ Myr. 
We also find that the majority of LBGs brighter than $M_V=-21.0$ 
have extended star formation activity at higher redshifts ($z=3-10$)
over a period of 1~Gyr, resulting in a median stellar mass
of $10^{10}\himsun$ at $z=3$, in accordance with 
the picture proposed by \citet{Steidel96}.
There seems to be no clear-cut choice between the two 
star formation scenarios: the bright LBGs in our simulation 
exhibit both instantaneous and continuous star formation activity.
Our result seems to be in qualitative agreement with the star 
formation histories inferred from recent near-IR observations of LBGs 
\citep{Papovich01, Shapley01}.
The stellar content and the merger history we observe in the 
simulation is not consistent with the picture that LBGs 
are low-mass starbursting objects which later evolve into low-mass 
spheroids at $z=0$ \citep{Lowenthal97, Sawicki98, Somerville01}.

We argued in Paper~I that the global star formation rate 
in the simulation increases toward higher redshifts up to $z\sim 8$.
Together with the results presented in this paper, the existence of 
high star formation rate galaxies at high-redshift seems to be 
a generic prediction of the CDM model. 
It is an important future task to investigate the link
between these galaxies in the simulation and the observation 
of high-redshift galaxies such as the Hubble Deep Field \citep{Williams96} 
and the future Next Generation Space Telescope \citep{Stockman96}.

Our results on the star formation rates of high-redshift
galaxies are in good agreement with the previous studies by 
\citet{Dave99} and \citet{Weinberg00}. This is not surprising
given that the simulations used in both studies adopt a similar 
star formation recipe, although the spatial and the mass resolutions 
are very different.

The sub-solar metallicity of LBGs at $z=3$ is naturally predicted  
by our simulation. The entire luminosity-metallicity distribution 
shifts to lower metallicity at higher redshifts. 
If indeed DLAs have metallicity of $\sim (1/30)\Zsun$ \citep{Pettini00}, 
then LBGs typically have higher metallicity than DLAs at $z=3$.
The connection between LBGs and DLAs in our simulation 
will be studied in a separate paper \citep{Cen01}.


\acknowledgments 
This paper is based on the thesis work of K.N. at Princeton University.
The author is grateful to Drs. Jeremiah P. Ostriker, Renyue Cen, 
and Masataka Fukugita for the collaboration on the two previous papers 
(Paper I \& II), and for advice, support and important scientific 
input to the present work. Drs. Renyue Cen and Jeremiah P. Ostriker 
are thanked for providing the simulation for the present work. 
We thank Dr. Chuck Steidel for useful discussions and wonderful
Spitzer Lecture Series given at the Peyton Hall of Princeton University 
in May 2001, which largely inspired this work. 
K.N. is grateful to Alice Shapley for making the draft of an 
unpublished paper available, and to Kurt Adelberger for the 
$\Un, G, \R$ filter functions. We thank Michael Strauss for useful 
comments on the draft. This work was supported in part by grants 
AST~98-03137 and ASC~97-40300. K.N. is supported in part by the 
Physics Department of Princeton University.



\clearpage
\begin{deluxetable}{cccccccc}  
\tablecolumns{8}  
\tablewidth{0pc}  
\tablecaption{Schechter Parameters in Rest-frame $V$ Band \label{table1}}  
\tablehead{
\colhead{Quantities} & \colhead{z=0} & \colhead{z=0.3} & \colhead{z=0.5} & 
\colhead{z=1} & \colhead{z=2} & \colhead{z=3} & \colhead{z=5}
}
\startdata
$\Mstarv$  & $-22.15$ & $-22.59$ & $-22.68 $ & $-22.64$ & $-22.50$ & $-22.05$ & $-21.17$\cr
$\phi^*$   & 0.83     & 0.90     & 0.97      & 1.17     & 1.70     & 2.70     & 5.50\cr
$\Ldenv$   & 1.73     & 2.82     & 3.29      & 3.82     & 4.89     & 5.10     & 4.66\cr 
\enddata
\tablecomments{
Parameters of the Schechter functions in rest-frame $V$ band before
taking dust extinction into account.
The faint-end slope $\alpha$ is fixed to $-1.15$ in all cases. 
See text for the method of fitting $\Mstarv$ [mag] and $\phi^*
[10^{-2}h^3\mpc^{-3}]$.
$\Ldenv$ is in units of [$10^8\Lsunv\mpc^{-3}$] with $h=0.67$.
Similar table for $B$ band was presented in Paper II.
}
\end{deluxetable}  


\clearpage

\clearpage

\clearpage

\clearpage

\clearpage

\clearpage

\clearpage

\clearpage

\clearpage

\clearpage

\clearpage


\end{document}